\documentclass[preprint,12pt]{elsarticle}

\usepackage{graphics}

\usepackage{amssymb}

\begin{document}

\begin{frontmatter}



\title{Statistical properties of agent-based models in markets with continuous double auction mechanism}


\author[as]{Jie-Jun Tseng}
\author[ntu]{Chih-Hao Lin}
\author[ntu]{Chih-Ting Lin}
\author[ncu]{Sun-Chong Wang}
\author[as]{Sai-Ping Li}

\address[as]{Institute of Physics, Academia Sinica, Nankang, Taipei 115, Taiwan}
\address[ntu]{Graduate Institute of Electronics Engineering, National Taiwan University, Taipei 106, Taiwan}
\address[ncu]{Institute of Systems Biology and Bioinformatics, National Central University, Chungli 320, Taiwan}

\begin{abstract}
Real world markets display power-law features in variables such as price fluctuations in stocks. 
To further understand market behavior, we have conducted a series of market experiments on
our web-based prediction market platform
which allows us to reconstruct transaction networks among traders. 
From these networks, we are able to record the degree of a trader, 
the size of a community of traders, 
the transaction time interval among traders and other variables that are of interest. 
The distributions of all these variables show power-law behavior. 
On the other hand, agent-based models have been proposed to study the properties of real financial markets. 
We here study the statistical properties of these agent-based models and compare them
with the results from our web-based market experiments. 
In this work, three agent-based models are studied, namely, zero-intelligence (ZI),
zero-intelligence-plus (ZIP) and Gjerstad-Dickhaut (GD). 
Computer simulations of variables based on these three agent-based models were carried out. 
We found that although being the most naive agent-based model,
ZI indeed best describes the properties observed in real markets. 
Our study suggests that the basic ingredient to produce the observed properties from
real world markets could in fact be the result of a continuously evolving dynamical
system with basic features similar to the ZI model.
\end{abstract}

\begin{keyword}
Econophysics \sep Financial stylized facts \sep Agent-based simulation \sep Complex networks \sep Continuous double action
%
\PACS 87.23.Ge \sep 89.65.Gh \sep 89.75.Da
\end{keyword}
\end{frontmatter}

%
\section{Introduction}
Many complex systems exhibit distributions of observables that have power-law behavior.
Examples include net wealth, earthquake magnitudes and gene expressions~\cite{powerlaw,bio}.
In economics, financial markets are complex systems that involve human activities and behavior.
Financial markets, consisting of such heterogeneous agents as arbitrageurs,
hedgers and investors, show stylized distributions of returns and wealth~\cite{4}.
The prices and individual wealth in these markets are driven up and down
by the so-called ``invisible hand" as coined by Adam Smith.
As a result, one will have time series for price and volume fluctuations.
Correlations of these quantities can be obtained and display interesting phenomena.
Intrigued by the universal behavior,
physicists have applied the methodologies of non-equilibrium statistical mechanics
to elucidate the mechanisms underlying the complexity~\cite{4}.
Examples include critical phenomenon~\cite{5} and self-organized criticality~\cite{6}
modeling of economic systems.
In an attempt to understand the features displayed in real markets, we set up a
web-based prediction market platform several years ago in order to monitor
the trading behavior among the human traders in real time.
Our platform, Taiwan Political Exchange (TAIPEX),
is a futures market with continuous double auction trading mechanism.
Beginning 2004, we have carried out several experiments on this platform and
have found many interesting features in these experimental markets~\cite{7a,7,8}.
Aside from the price and volume time series which resemble real market data,
we have indeed also found power-law behavior in the degree distribution of
the traders' transaction network, the wealth distribution of traders and community size,
etc some of which have never been able to be extracted from real markets before.
Together with the information obtained from experiments on our platform and other real markets,
one should be able to perform more detailed analysis that could lead to a better understanding
of trading behavior among human traders.
In economics, double auction is one of the most widely used mechanisms in all kinds of markets
including stock exchanges and business-to-business e-commerce.
The convergence and efficiency properties of the double auction institution
has also been the subject of intense interest among experimental economists,
beginning with the work of Smith~\cite{9}, who built on the early work of Chamberlin~\cite{10}.
In experimental economics, people have in recent years designed computer-based agent models
to study these properties.
Their main concern is to investigate the convergence and efficiency properties
of the double auction mechanism in real financial markets.
With this new set of data from our platform and tools borrowed from physics in hand,
it is natural to ask if they can put further constraints on agent-based model building in economics.
 We will make such an attempt here.
We choose here three agent-based models, namely, zero-intelligence (ZI)~\cite{11},
zero-intelligence-plus (ZIP)~\cite{12} and Gjerstad-Dickhaut (GD)~\cite{13}.
These three agent-based models are commonly used in economics to study
the market behavior and are the bases of many other more complicated agent-based models.
We will perform Monte Carlo simulations in these models and compare the simulated results
of the agent-based models to that of the results from our market experiments.
In section~\ref{sec:abm}, we introduce the basic ingredients of the three agent-based models
that are investigated in this paper.
Section~\ref{sec:result} contains results from our simulation using these models
and we compare them with the results from TAIPEX experiments in section~\ref{sec:comparison}.
Finally, we summarize our findings in section~\ref{sec:summary}.
\section{Markets with agent-based models}
\label{sec:abm}
With the advent of faster and cheaper computing power,
agent-based models are being employed to study phenomena in economic systems such as financial markets.
One of the earliest such models is called the ``zero-intelligence" agent model proposed by
Gode and Sunder~\cite{11} back in 1993.
The ZI traders are, by definition, agent traders without any intelligence.
In the markets, they will submit random bids and offers.
Therefore the resulting price never converges toward any specific level.
There are many variations of the ZI model and we here make two choices to simplify our simulation.
First, during each bid, offer and transaction are valid for a single item.
Second, in each duration, every trader could make only one successful transaction
(i.e., the buyer can only have one item to buy and the seller only has one item
to sell in each duration).
The implementation of our simulation is as follows: For the structure of markets,
the supply and demand functions are generated from Smith¡¦s value mechanism~\cite{14}
at the beginning for each run and will not change throughout the simulation.
There are an initial fixed number of ZI traders in our simulation.
Half of them are classified as buyers and the remaining half of the traders are sellers.
At each step, one buyer and one seller are chosen for the matching.
Due to the budget constraint,
the buyer must bid with the price lower than its redemption value given by
the demand function and the seller must offer the commodity
at the price higher than the cost generated by the supply function.
Once the bidding price exceeds the offering price,
the transaction between this buyer and seller will be made.
No transaction will be made otherwise.
Whether a successful transaction occurs or not,
the system will move forward to the next step and choose another pair of traders.
The simulation lasts for a period of p sessions (days), each having d rounds.
The simulation will therefore terminate after $p\times d$ steps.

Since then, many agent-based models have been proposed, with various degrees of complication.
Among the various models are two popular models, the ZIP~\cite{12} and the GD~\cite{13} models.
These models are designed to have a better convergence of the price to its equilibrium value.
The ZIP model can be viewed as a modified version of ZI.
Similar to the ZI traders, these simple agents make stochastic bids.
In addition, the ZIP agents employ an elementary form of machine learning.
The learning mechanism here depends on four factors.
The first factor is whether the trader is active or inactive.
The other three factors all concern the last (or most recent) event: the price,
whether it was a bid or an offer and whether a transaction was made or not.

In the case of GD, each buyer forms a subjective belief that some seller
will accept his bid and determines which bid will maximize his expected profit.
In a similar way, each seller forms a subjective belief that some buyer
will accept his offer and determines his offer in order to maximize his expected profit.
These beliefs are based on the observed market data including frequencies of asks,
bids, accepted asks, accepted bids, etc.
More details of the ZIP and GD models can be found in~\cite{12,13}.
\section{Simulation results}
\label{sec:result}
\begin{figure}
  \centering\includegraphics[width=0.5\textwidth]{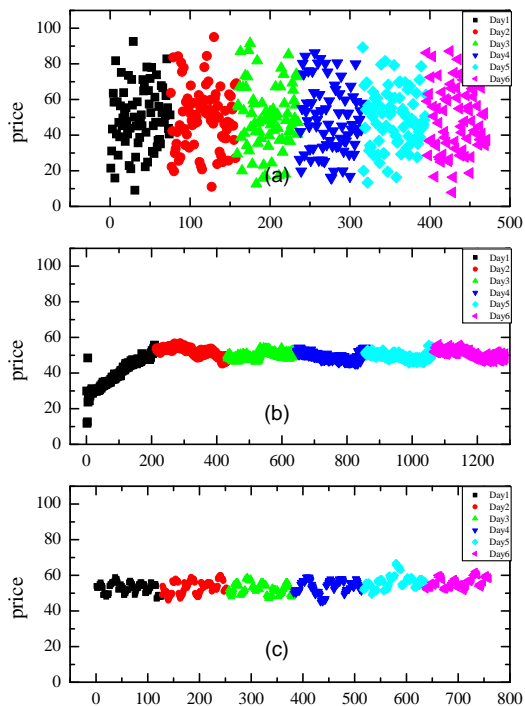}
  \caption{Price time series of (a) ZI, (b) ZIP and, (c) GD for the first 6 days of our simulations.}
  \label{fig1}
\end{figure}

Simulations using the three agent-based models (ZI, ZIP and GD) were carried out.
In order to compare with results from the prediction markets on our platform and also
real financial markets, we performed simulations on the degree of a trader
(the number of traders that a trader has transactions with), the size of a community of traders
(who have similar trading behavior),
the transaction time interval among traders and also the price fluctuations.
In all simulations below, we set the supply and demand curves to take values
between 0 and 100 with the buyers and sellers randomly distributed on the two curves.
To be more specific, we use straight lines for the supply and demand curves
and the buyers and sellers fall randomly on these lines.
Other curves with reasonable shapes can be used but do not affect our conclusion below.
Unless otherwise stated, we set $N$, the number of agents to be 2500.
On each transaction day, we performed 2000 rounds.
One round here means we randomly picked one buyer and one seller and checked
whether the price could match.
If they matched, a transaction was said to be made.
We did this for a total of 200 days.
The results are presented below.
Ten runs were taken and averaged in each of the cases studied.
The price time series of the three models are shown in Fig.~\ref{fig1} for the first 6 days
of our simulations.
As expected, the price time series of market with ZI agents exhibits continuous large fluctuations
while ZIP and GD tend to converge to the equilibrium price value. 
%
\subsection{Degree distribution}
\begin{figure}
  \centering\includegraphics[width=0.5\textwidth]{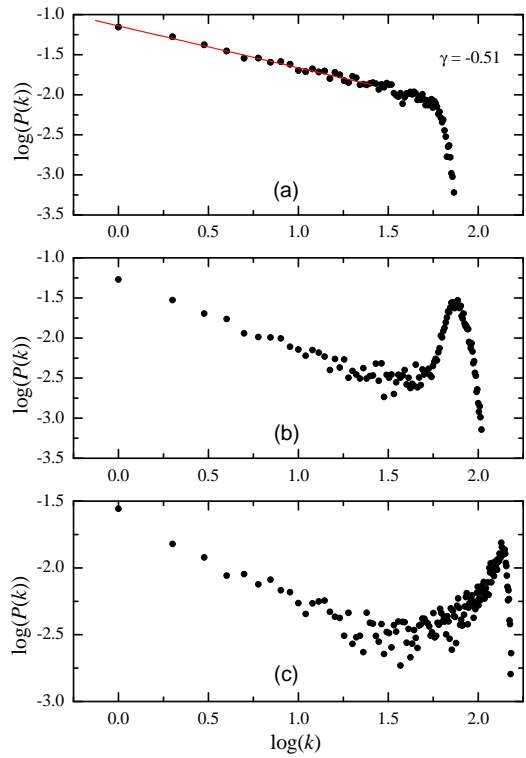}
  \caption{The degree distribution of the transaction networks in three agent-based models from simulation:
  (a) ZI, (b) ZIP and, (c) GD.}
  \label{fig2}
\end{figure}
While it is impossible to obtain data about the transactions
among individual accounts in real markets,
it is possible to record such activities in the experimental markets conducted on our platform~\cite{7}.
One can therefore be able to build transaction networks among traders and study these complex networks.
In a transaction network, a trader is denoted by a node.
A link between two nodes in the transaction network therefore indicates that
there is at least one transaction between the two traders.
The degree of a trader (node) therefore means the number of links of a trader (node)
to other traders (nodes).
One should notice that, although the total number of traders is fixed at the beginning,
not all of them will make a successful transaction with others.
The final number of nodes connecting to the whole network
(i.e., traders with successful transactions),
the total number of links and the average degree distribution
$\langle k \rangle$ will depend on the input values of rounds and days.
Fig.~\ref{fig2} illustrates the result of the degree distribution from the three agent-based models.
%
\subsection{Community size distribution}
\begin{figure}
  \centering\includegraphics[width=0.5\textwidth]{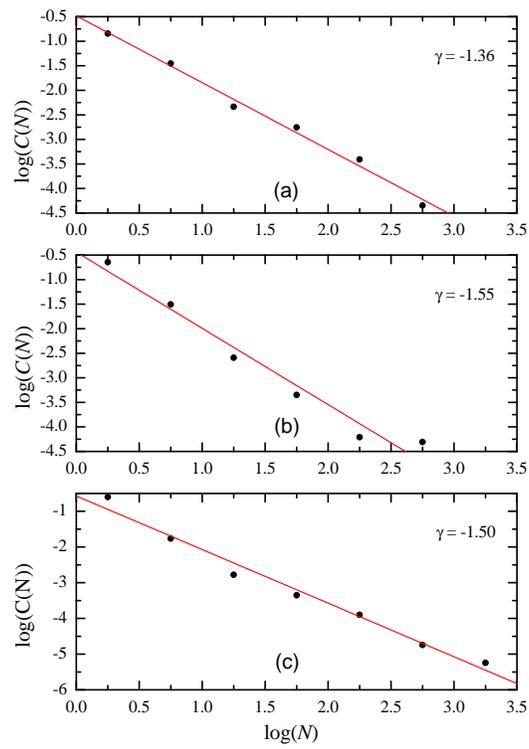}
  \caption{Distributions of the community sizes from the transaction networks of
  (a) ZI, (b) ZIP and, (c) GD agents.}
  \label{fig3}
\end{figure}

In addition to the degree distribution,
we also study the community size distribution of the transaction networks.
In the context of the market experiments performed on our platform~\cite{7},
when the price of a futures contract was considered too high (low), a sell (buy) order was placed.
A link between two nodes in the transaction network therefore indicates that the
two traders disagreed on the pricing of the futures contract.
In other words, traders with no links between them were those who thought alike.
An algorithm to find communities of the traders is thus to partition the transaction network
such that the densities of edges within communities are lower and those between communities
are higher than average.
We here applied the eigenvector-based partitioning algorithm of~\cite{15}
to the networks from the simulation of the three agent models and the result is shown in Fig.~\ref{fig3}.
All three models display power-law behavior with exponents $-1.36$ (ZI), $-1.55$ (ZIP) and $-1.50$ (GD).
%
\subsection{Transaction time interval}
\begin{figure}
  \centering\includegraphics[width=0.5\textwidth]{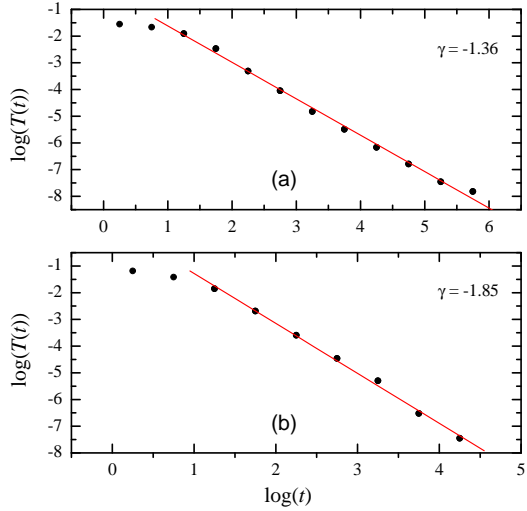}
  \caption{Distribution of time intervals between successive transactions of (a) ZI and, (b) ZIP.}
  \label{fig4}
\end{figure}

While one can record the transaction time interval between transactions in real markets,
it is not obvious how to implement this in agent-based model simulations.
The basic problem is how to relate the real market calendar time to the Monte Carlo simulation
in agent-based models.  Each of the Monte Carlo steps might not correspond
to the same amount of time in real market.  
We here define the transaction time interval in the simulations to be the Monte Carlo steps
between two transactions.
Since our aim here is to study the time interval between transactions,
we instead make each run to have 1 million rounds and we do it for 30 times.
This is to eliminate finite size effect and also to resemble the experiments run on our platform.
We further note the case of GD, each trader is given a probability to be selected for a transaction.  
When two agents are picked in each Monte Carlo step, it will always result in a transaction.
There is no transaction time interval between each transaction and thus no transaction 
time interval can be defined for GD.
The result of our simulation on ZI and ZIP are shown in Fig.~\ref{fig4}.
%
\subsection{Price fluctuation}
\begin{figure}
  \centering\includegraphics[width=0.5\textwidth]{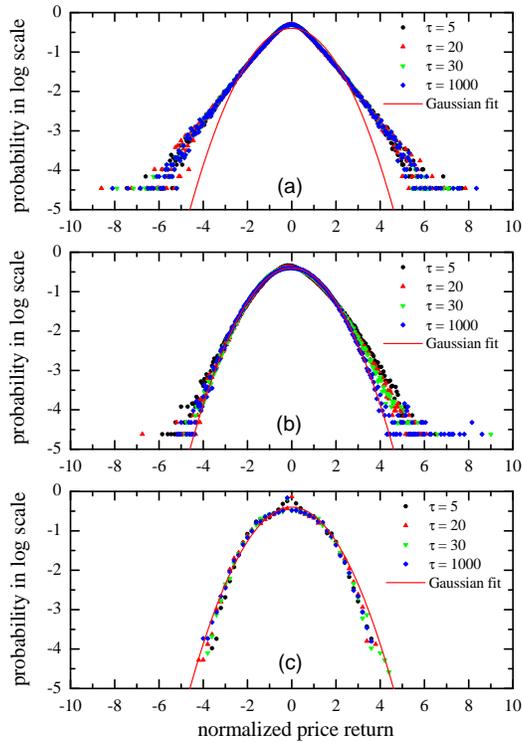}
  \caption{Probability density of normalized price return with different time lags $\tau$:
  (a) ZI, (b) ZIP and, (c) GD. The time lags in all cases are 5, 20, 30 and 1000.
  Red lines are Gaussian fits to the data.}
  \label{fig5}
\end{figure}

In real markets, there always appear some large fluctuations in stock prices. 
To study occurrence of the fluctuations,
we calculate the difference in the logarithmic price $\log S(t)$ between time $t+\tau$ and time $t$,
\begin{equation}
  G_{\tau}(t) = \log S(t+\tau) - \log S(t);
\end{equation}
and the normalized price return $g_{\tau}(t)$,
\begin{equation}
  g_{\tau}(t) = \frac{G_{\tau}(t)-\mu_{\tau}}{\sigma_{\tau}};
\end{equation}
where $\mu_{\tau}$ and $\sigma_{\tau}$ are the mean and standard deviation of $G_{\tau}(t)$. 
Price fluctuations can be studied by two approaches.
The first is the price fluctuations in real time and the second is the price fluctuations
between consecutive transactions~\cite{16,17}.
Since it is not obvious how to relate the Monte Carlo time steps to calendar time,
we prefer to follow the second approach.
We therefore define the time t to be the sequence of successful transactions and each time step
refers to one successful transaction here.
The result is shown in Fig.~\ref{fig5}.
%
\section{Comparison with TAIPEX results}
\label{sec:comparison}
\begin{figure}
  \centering\includegraphics[width=0.5\textwidth]{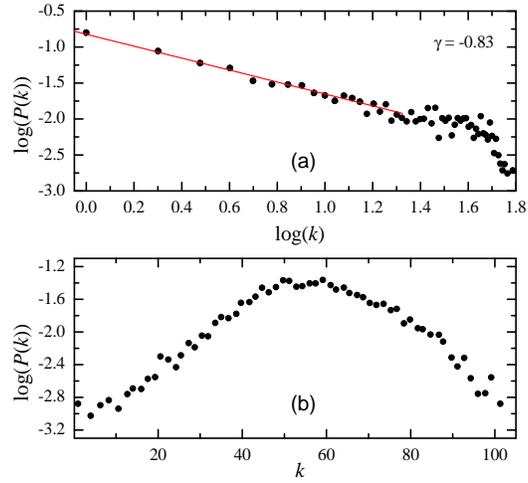}
  \caption{Degree distribution of (a) extramarginal traders and, (b) intramarginal traders in the ZIP simulation.}
  \label{fig6}
\end{figure}

\begin{figure}
  \centering\includegraphics[width=0.5\textwidth]{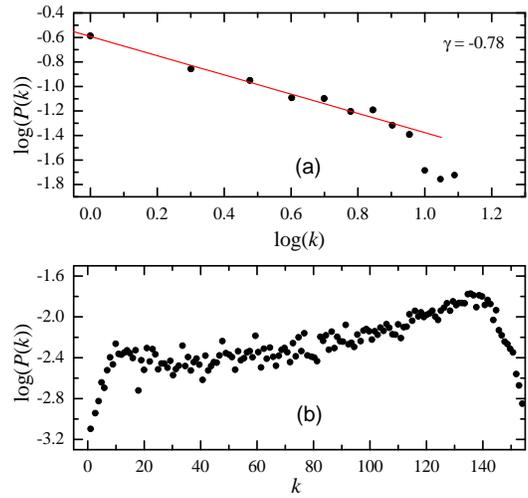}
  \caption{Degree distribution of (a) extramarginal traders and, (b) intramarginal traders in the GD simulation.}
  \label{fig7}
\end{figure}

\begin{figure}
  \centering\includegraphics[width=0.5\textwidth]{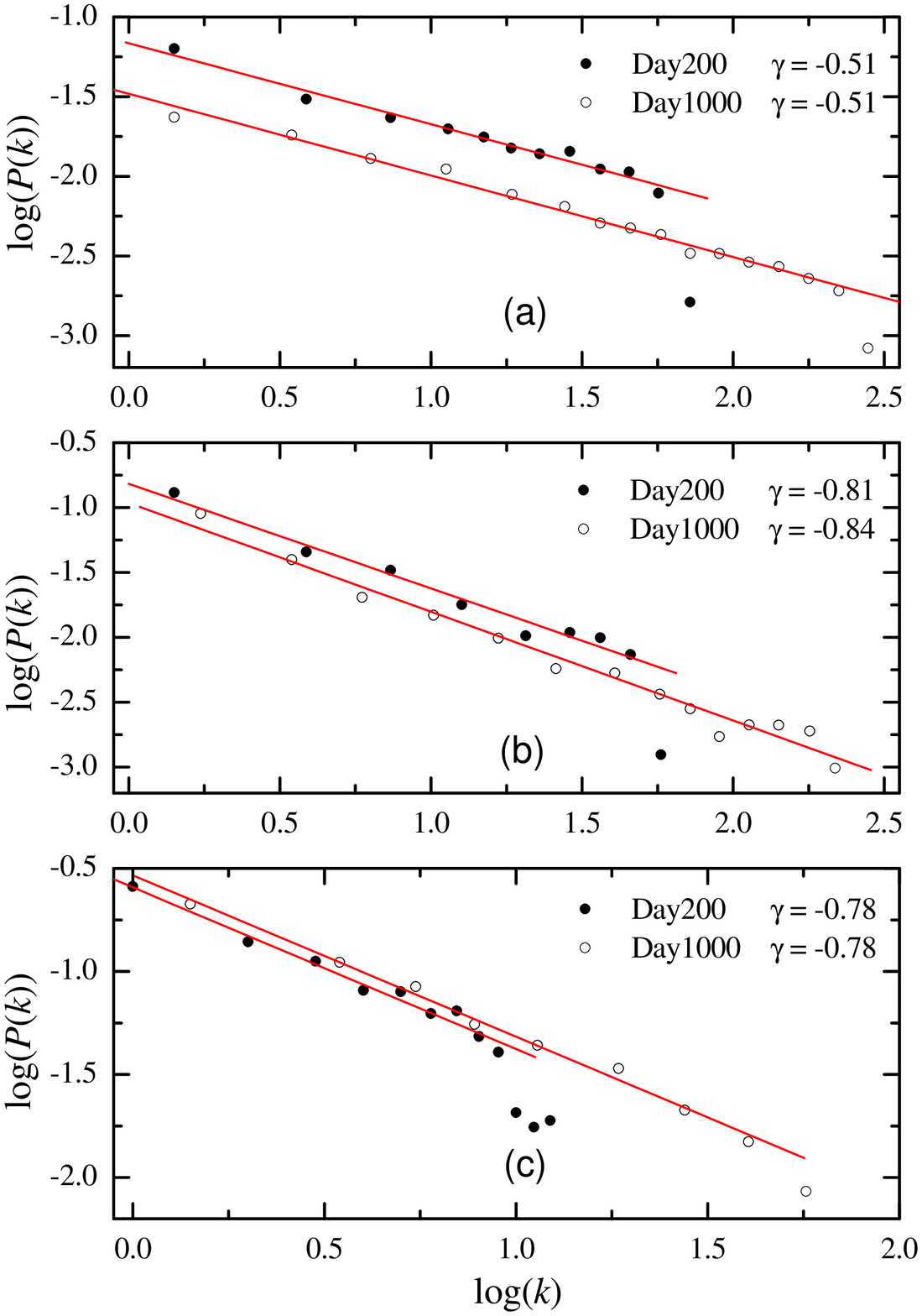}
  \caption{Degree distribution of (a) ZI agents,
  (b) ZIP agents and, (c) GD agents from 200 and 1000 day period simulations}
  \label{fig8}
\end{figure}

In the above, we presented the results of our simulations based on ZI, ZIP and GD agents.
In the study of degree distribution, our result shows that the degree distribution of ZI agents
has a power-law behavior with an exponent of about -0.51 and approaches zero
quickly near the tail, which we believe is due to finite size effect and will be investigated below.
The average degree distribution $\langle k \rangle$ in this case is about 20.
In the case of ZIP and GD agents, however, we observe that there is a bump at the tail in each of the models.
To understand how this comes about, we further separate the traders into two sectors,
namely the intramarginal traders and the extramarginal traders.
For those to the left-hand side of the intersection of supply and demand curves are defined as intramarginal traders,
while those to the right-hand side of the intersection are extramarginal traders.
The analysis here suggests that the bump in each case is a result of the fact that
after an equilibrium price is reached, only intramarginal traders can contribute to transactions made.
This means that a transaction will be made only if we pick a pair of intramarginal buyer and seller.
The bump is therefore a result of the contribution from intramarginal traders.
In contrast, extramarginal traders can never contribute to transactions made after the equilibrium price is reached.
The degree distributions resulted from ZIP and GD agents
are replotted in Fig.~\ref{fig6} and Fig.~\ref{fig7}, respectively.
One can see that in the case of ZIP agents, degree distribution of the extramarginal traders (Fig.~\ref{fig6}(a))
follow an approximate power-law with an exponent of about $-0.83$
while that of the intramarginal traders (Fig.~\ref{fig6}(b)) is close to a normal distribution.
In the case of GD agents (Fig.~\ref{fig7}(a)),
it is not obvious if the degree distribution of the extramarginal traders also follows
a power-law behavior since it seems that finite size effect dominates
when we perform runs with a 200 day period. 
In order to clarify this issue, we have carried out simulations for longer periods.
A 1000 day period simulation was done and is shown in Fig.~\ref{fig8}.
A curve with power-law behavior appears with an exponent equal to $-0.78$ for GD agents in Fig.~\ref{fig8}(c).
For comparison, we also include here
the 1000 day period simulation with ZI and ZIP agents in Fig.~\ref{fig8}(a) and (b), respectively.
The result supports our belief that the drop at the tails is due to the finite size effect.
We therefore conclude that all three models exhibit power-law behavior
in the degree distribution of transaction networks, with exponents ranging $-0.84 \sim -0.51$.
As a comparison, we note that results from our TAIPEX platform gives an exponent
of about $-2.14$ in the latest experiment conducted in March 2008~\cite{8},
from a transaction network of 1985 traders.
Although these agent-based models exhibit power-law behavior in the degree distribution of transaction networks,
the exponents are very different from those obtained from our market experiments with human agents.
In the case of community size distribution, all three models studied here show power-law
behavior with exponents ranging $-1.36 \sim -1.55$.
This is close to the result obtained from TAIPEX~\cite{7}, which is about $-1.2$.
The transaction time interval of ZI and ZIP agents as defined also show power-law behavior
with exponent $-1.36$ and $-1.84$ respectively.
As a comparison, the result from TAIPEX is about $-1.3$.
Although one cannot directly compare the exponents between real markets
(such as those obtained from our market experiments) and the agent-based models,
it is still interesting to know that transaction time intervals in agent-based models
exhibit such power-law behavior.

The results of price fluctuations were presented in Fig.~\ref{fig5}.
Curves of different time lags ($\tau$ = 5, 20, 30, 1000) in each case were plotted.
Each of the three models shows interesting features different from each other.
One can see that of the three models studied here, only market with ZI agents shows a significant difference
from a Gaussian distribution.
Furthermore, the curves with different time lags in Fig.~\ref{fig5}(a) fall onto the same curve,
indicating scaling behavior.
In the case of ZIP agents, the curves of different time lags do not fall on the same curve,
with $\tau$ = 5 somewhat away from a Gaussian distribution while $\tau$ = 1000 follows
such a Gaussian distribution.
We should remind our readers that
since the time lags here are referred to the number of transactions in between,
our result might only be viewed as an indication
that ZI agents should better describe the price fluctuations in real markets.
\section{Summary}
\label{sec:summary}
Our results illustrate that although being the most naive agent-based model,
ZI agents indeed capture the above mentioned features in real markets.
A feature which distinguishes ZI from the other two agent-based models is that
while the market values of the assets in these models all tend to converge to an equilibrium value,
continuous large fluctuations are observed in ZI.
It is known that in real financial markets,
there is an excess of large fluctuations compared to a Gaussian distribution,
which probably results from the time-varying volatility.
Therefore, one will see heavy tails in a plot of the probability density of normalized price return
with different time lags,
which are significantly different from Gaussian distributions,
similar to Fig.~\ref{fig5}(a) but with even heavier tails.
Any agent model should at least capture this generic feature.
It is easy to see that of the three models, only ZI agents can capture this feature.
It could probably imply that the market with ZI agents has a time evolving volatility
which can reproduce the features in a real financial market.
Although the exponents of various parameters in ZI are different from real markets,
it nevertheless is the first step toward building a better agent model to mimic real financial markets.  
\section*{Acknowledgments}
This work was supported in part by the National Science Council of Taiwan under grants
NSC\#98-2120-M-001-002 and NSC\#97-2112-M-001-008-MY3.
\bibliographystyle{elsarticle-num}
\bibliography{li_paper}
\end{document}